# Modeling the Dynamics of Phospholipids in the Fluid Phase of Liposomes

Sudipta Gupta,*[a] and Gerald J. Schneider*[a,b]

[a]Department of Chemistry, Louisiana State University, Baton Rouge, LA 70803, USA
[b]Department of Physics & Astronomy, Louisiana State University, Baton Rouge, LA 70803, USA

**ABSTRACT**

We present the derivation of a new model to describe neutron spin echo spectroscopy and quasi-elastic neutron scattering data on liposomes. We compare the new model with existing approaches and benchmark it with experimental data. The analysis indicates the importance of including all major contributions into modeling of the intermediate scattering function. Simultaneous analysis of the experimental data on lipids with full contrast and tail contrast matched samples, reveals highly confined lipid tail motion. A comparison of their dynamics demonstrates the statistical independence of tail-motion and height-height correlation of the membrane. A more detailed analysis indicates that lipid tails are subject to relaxations in a potential with cylindrical symmetry, in addition to the undulation and diffusive motion of the liposome. Despite substantial differences in the chemistry of the fatty acid tails, the observation indicates a universal behavior. The analysis of partially deuterated systems confirms the strong contribution of the lipid tail to the intermediate scattering function. Within the time range from 5 to 100 ns, the intermediate scattering function can be described by the height-height correlation function. The existence of the fast-localized tail motion and the contribution of slow translational diffusion of liposomes determines the intermediate scattering function for $t < 5$ ns and $t > 100$ ns, respectively. Taking into account the limited time window lowers the bending moduli by a factor of 1.3 (DOPC) to 2 (DMPC) compared to the full range.

## 1 INTRODUCTION

Phospholipids are an essential part of cell membranes. Many recent studies focus on lipids and their impact on the proper functioning of membrane proteins.[1, 2] Nuclear magnetic resonance (NMR) is frequently utilized to explore the molecular dynamics of liposomes.[3] NMR reveals that lipid rotational and lateral motions were observed along with slow flip-flop motion where lipid exchange across the two monolayers.[3] Rotational diffusion of lipids plays an important role in transport of proteins, whereas,[4] lipid flip-flop motion is important for maintaining the stability and composition of the inner and outer monolayers of the membranes.[5] At length scale of the membrane thickness the entire membrane can undergo out-of-plane thickness and bending fluctuations or undulations.[6-8] Such motions are responsible for cellular uptake or release and pore formations in membranes.[9, 10] The size of liposomes is important for bio-engineering and reported in drug-deliver studies.[11] The diameter of liposomes marks the larger length scale and relates to the translational diffusion, $D_t$. So, from both theoretical and practical point of view it is important to have a universal model that can relate different dynamics over multiple length and time scales.

The connection between the hydrodynamic size and diffusion via the Stokes-Einstein equation makes dynamic light scattering (DLS) a well-established tool to determine the translational diffusion coefficient, size and size distribution of liposomes.[12] Microscopic techniques at larger lipid domains, e.g., fluorescence recovery after photobleaching (FRAP)[13] and single-particle tracking (SPT)[14, 15] with fluorescent labelling can be utilized to determine the lateral diffusion coefficient and mean squared displacement of lipids. Compared to neutron spectroscopy, fluorescent labelling techniques generally probes larger length scales and are limited by their temporal resolution. In addition, they may require a fluorescence dye that may lead to additional effects, especially when tracking particle trajectories.[16][17] More importantly, due to their fast motion at the ps to sub-µs time scale, studying the dynamics of fatty acid tails is impossible by microscopy and outside the length scale window of DLS.

Several, non-invasive neutron scattering techniques exist that are very useful to explore the structure and dynamics at the appropriate length and time scales of the living cells in their natural state.[8,18] Structural details can be obtained by selective deuteration and contrast variation.[19] Due to their importance, thickness fluctuations at the intermediate length scale have been extensively studied by neutron spin echo spectroscopy (NSE). [6,19,20]

In this context, the time-dependent mean-squared displacement (MSD or $\langle \Delta r^2(t) \rangle$) is one of the most fundamental means of statistical physics to describe the molecular dynamics of a molecule or the ensemble average. Since the MSD provides valuable information it is often used to track molecular motions or changes due to the influence of interactions and spatial confinements in crowded biomacromolecules and polymers.[15,21-24] Recently, we utilized NSE to explore the MSD of lipids at the time scale around 50 ps to 200 ns.[25,26] We compared four different phospholipid samples, DOPC (1,2-dioleoyl-sn-glycero-3-phosphocholine), DSPC (1,2-distearoyl-sn-glycero-3-phosphocholine), DMPC (1,2-dimyristoyl-sn-glycero-3-phosphocholine) and SoyPC (L-α-phosphatidylcholine), in their fluid phases.[27]

By a detailed calculation of the time evolution of $\langle \Delta r(t)^2 \rangle$, we obtained three distinct power-laws in the time range of the NSE experiment. We found $t^1$ at longer Fourier times, followed by $t^{0.66}$ and $t^{0.26}$ ($t < 5$ ns), at intermediate and shorter Fourier times, respectively. The $t^1$ ($t > 80$ ns) contribution relates to the center of mass diffusion of the liposomes, whereas the $t^{0.66}$ (5 ns $<$ $t$ $<$ 80 ns) originates from the thermal undulations of the membrane as defined by Zilman-Granek (ZG),[28] and also by the anomalous diffusion predicted by Monte Carlo simulations.[29] A power-law dependence of the specific strength of interactions was proposed by Pandey et al.[29], ranging from 0.17 (ΔF° $>$ 0) to 0.34 (ΔF° $<$ 0), with, ΔF°, the change in membrane-membrane interaction energy. Recent Molecular Dynamics (MD) simulations and mode-coupling theory calculations by Flenner et al.[30] relate trapped motion with the dynamics of the lipid tail of the fatty acid.

According to the simulations, the existence of anomalous diffusion seems to coincide with increasing disorder of the lipids, e.g., due to increase in temperature or addition of cholesterol.[31] Similar observations were reported for natural membranes where proteins are present to transport ions or genetic code across the membrane.[32] In such crowded environments, significant inhomogeneities were observed in single-particle trajectories, resulting in non-Gaussian diffusion.[32]

Neutron spectroscopy measures the spatial and temporal correlation functions simultaneously, with the additional advantage of the isotopic selectivity. Hereafter, we show the derivation of a constitutive model that describes all processes identified in the time- and length scale region of the NSE experiment. For the sake of completeness, we have discussed our model in relationship with models from literature and have compared results. This discussion is important because it reveals which cases require the new model. Hereafter, we start with a derivation of the new model and a comparison with existing models from the literature. We continue with a comparison of experimental results partly taken from the literature.

## 2 Basics
**Cumulant approach**

Within the framework of Gaussian approximation, the intermediate scattering function $S(Q,t)$ as obtained from NSE, and the mean squared displacement $\langle \Delta r(t)^2 \rangle$ are related by

$$\frac{S(Q,t)}{S(Q)} = A \exp\left[-\frac{Q^2 \langle \Delta r(t)^2 \rangle}{6}\right] \quad (1)$$

For a more generic case, $S(Q,t)$ can be expressed by a cumulant expansion [33][34][35]

$$\frac{S(Q,t)}{S(Q)} = A \exp\left[-\frac{Q^2 \langle \Delta r(t)^2 \rangle}{6} + \frac{Q^4 \alpha_2(t)}{72} \langle \Delta r(t)^2 \rangle^2\right] \quad (2)$$

The last equation introduces the non-Gaussianity parameter, $\alpha_2(t) = \frac{d}{d+2} \frac{\langle \Delta r(t)^4 \rangle}{\langle \Delta r(t)^2 \rangle^2} - 1 = \frac{d}{d+2} \beta_2 - 1$. The parameter, $\alpha_2$, is a very convenient means to indicate deviations from the often assumed Gaussian approximation.[S4,S5] The kurtosis, $\beta_2$, is defined by the quotient of the fourth $\langle \Delta r(t)^4 \rangle$ and the second moment squared $\langle \Delta r(t)^2 \rangle^2$. In this paper, we have introduced a generalized approach and explored the limit $Q \to 0$ to understand the overall mean squared displacement (MSD).

## 3 RESULTS and DISCUSSION

This section shows the derivation of a new model to describe the dynamics of liposomes as measured by neutron spectroscopy and discusses the differences in relationship with existing models, and is then benchmarked against the experimentally determined intermediate scattering function, $S(Q,t)$.

### 3.1 Derivation of a new model

In order to derive our new model, we must consider different contributions to the dynamics of liposomes as reported from different experiments. The data demands a generalized approach that includes translational diffusion of the liposomes, and collective fluctuations of the membrane. Taking this into consideration we present step by step derivation of the unified model.

### 3.1.1 Separation ansatz, statistical independence of different contribution to intermediate scattering function

Based on our recent paper,[25] we know that the translational motion of the liposome is independent of the lipid motion, at least within a very good approximation. The experiments show at least three processes, tail motion, collective lipid motion of the membrane and translational diffusion of liposome that contribute to the time-dependent mean squared displacement (MSD or $\langle r^2(t) \rangle$) within the length and time window of the NSE.

Using partially deuterated lipids where the lipid tail is contrast matched with the solvent,[6] it is evident that the height-height correlation function can be well described by the Zilman-Granek (ZG) approximation for membrane undulation.[28] The ZG approximation neglects the contribution of the lateral and more local motions of lipids to $S(Q,t)$.

Our experiments presented below demonstrate that the timescales are well separated, and the fast-local relaxation of lipids and the height-height correlation of membranes can be treated as statistically independent contributions. Therefore, we assume the faster lipid tail motion is not affected by the slower ZG dynamics.[25] Hence, the intermediate scattering function of the liposome, $S_{liposome}(Q,t)$, can be written as

$$S_{liposome}(Q,t) = S_{tail}(Q,t) \times S_{height}(Q,t) \times S_{thickness}(Q,t) \times S_{trans}(Q,t) \quad (3)$$

Here, the lipid bilayer motion is given by the height-height correlation of the membrane represented by $S_{height}(Q,t)$ (ZG), and the bilayer thickness fluctuation, $S_{thickness}(Q,t)$. The localized motion of the lipid tail in the bilayer is introduced by $S_{tail}(Q,t)$, whereas the translational motion of the liposome is given by $S_{trans}(Q,t)$. Following the literature, e.g., the textbook of Higgins and Benoit,[36] this approach is strictly valid if the different motions are statistically independent. Our experiments suggest that this assumption should be fulfilled at least to a very good approximation.

Equation 3 permits to include multiple processes, including rotational diffusion of liposomes and lipids. These processes are beyond the scope of the present work.

### 3.1.2 Contribution of diffusive motion

The diffusive motion of liposomes can be expressed as a function of time using the momentum transfer, $Q$, and the translational diffusion coefficient, $D_t$, as:

$$S_{translation}(Q,t) = \exp(-D_t Q^2 t) \tag{4}$$

Zilman-Granek discuss the impact of translational diffusion on the intermediate scattering function and introduce $D \sim k_B T/\eta R$, with the thermal energy $k_B T$ compared with the product of the viscosity, $\eta$, and the size of the liposome, $R$. They mention for $QR \gg 1$ the contribution of diffusion on $S(Q,t)$ is negligible for $t \ll \eta R^3/\kappa$. This discussion includes that the relaxation of the intermediate scattering function $S(Q,t)$ diminishes to vanishingly small value for $t \gtrsim \eta R^3/\kappa$, which could make the contribution of the diffusion barely visible. As suggested by Zilman-Granek we have replaced the plaquettes size $\xi$ by the liposome radius $R$.[28]

### 3.1.3 Contribution of height-height correlation, Zilman-Granek model

The height-height correlation function describing the membrane undulation has been derived by Zilman and Granek and has been extensively tested in the literature[28]. Most studies use

$$S_{height}(Q,t) = A\exp\left[-\left(\Gamma_Q t\right)^{2/3}\right] \tag{5}$$

The parameter $\Gamma_Q$ or $\Gamma_{ZG}$ introduces a $Q$-dependent decay rate, from which we derive the intrinsic bending modulus, $\kappa_\eta$, by[7, 37, 38]

$$\frac{\Gamma_Q}{Q^3} = \frac{\Gamma_{ZG}}{Q^3} = 0.0069\gamma\frac{k_B T}{\eta}\sqrt{\frac{k_B T}{\kappa_\eta}} \tag{6}$$

Here $\eta$ is the viscosity, $k_B$ the Boltzmann constant, $T$ the temperature, and $\gamma$ is a weak, monotonously increasing function of $\kappa_\eta/k_B T$.[28] In case of lipid bilayers, $\gamma$ has been found to be independent of $\kappa_\eta/k_B T$. Thus, the respective literature defines $\gamma = 1$.[6, 7, 28, 37, 39] This relationship is strictly valid for $\kappa_\eta/k_B T \gg 1$.[6, 7, 28, 37, 39] The numerical prefactor of 0.0069 seems to be the most up to date value as discussed in our recent review.[27] In **Table 3** we summarize $\kappa_\eta$ values from the literature. During years, literature has used different numerical prefactors. Therefore, to refer to 0.0069 the data from literature has been partly recalculated to avoid artificial differences. According to the Zilman-Granek the Stokes-Einstein diffusion coefficient of a single membrane plaquette of size, $r \sim Q^{-1} \left(\frac{\kappa_\eta}{k_B T}\right)^{1/2}$, can be written as, $D_{\text{eff}} \sim \frac{k_B T}{\eta} \left(\frac{k_B T}{\kappa_\eta}\right)^{1/2} Q$.[28] This determines the effective diffusion for the membrane undulation.

Following the work of ZG allows to introduce a relationship between the MSD (cf. ESI) and bending rigidity,

$$\frac{\kappa_\eta}{k_B T} = \frac{t^2}{c(\eta,T)^3 \langle \Delta r(t)^2 \rangle^3} \qquad (7)$$

with $c(\eta,T) = \frac{1}{6}\left(\frac{\eta}{0.0069 k_B T}\right)^{2/3}$. Equation 7 can be immediately obtained from the comparison of equations 5, 6, and 1. The comparison with the cumulant expansion (2) directly reflects the Gaussian assumption ($\alpha_2 = 0$) made by Zilman-Granek to derive their model. Hereafter we utilize the fact that within the framework of ZG model, $\langle \Delta r(t)^2 \rangle \propto t^{2/3}$. Consequently, displaying the bending rigidity as a function of time should yield, $\kappa_\eta/k_B T \propto t^2/t^2 =$ const.

### 3.1.4 Contribution of thickness fluctuations, Nagao model

Bilayer thickness fluctuations where monitored more in detail by NSE utilizing contrast matched fatty acid tails by Nagao and coworkers.[6, 37] The authors added an empirical Lorentzian function to equation 7, to account for the additional peak in the experimental data[6, 37]

$$\frac{\Gamma_Q}{Q^3} = \frac{\Gamma_{ZG}}{Q^3} + \frac{(\tau_{TF} Q_0^3)^{-1}}{1 - (Q - Q_0)^2 \xi^2} \qquad (8)$$

Where $\tau_{TF}$ is the relaxation time, and, $\xi^{-1}$ is the half width at half maximum of the Lorentzian at the thickness fluctuation peak momentum transfer, $Q_0$. To relate the observations with the physical properties Nagao et al. used a theoretical relation between thickness fluctuation and viscoelasticity of membranes derived by Bingham *et al.*[40]

By inserting equation 8 in equation 5 we obtain summation over two contributions, height correlation and thickness fluctuations. Therefore, equation 5 can be divided into the product of two contributions, $S_{height}(Q,t) \times S_{thickness}(Q,t)$. The separation into height and thickness correlations is mathematically equivalent to the factorization approach of equation 3, except the inclusion of localized fluctuation and translational diffusion of the liposome. More details about the thickness fluctuations are beyond the scope of the present work and can be found in the recent publication by Nagao *et al.*[37]

### 3.1.5 Contribution of confined motion of tails

The confined motion of the lipid tail can be described by

$$S_{tail}(Q,t) = \left( n_{H,head} + n_{H,tail} \left( \mathcal{A}(Q) + (1 - \mathcal{A}(Q)) \exp\left(-\left(\frac{t}{\tau}\right)^{\beta}\right) \right) \right) \quad (9)$$

with the relative number of protons in the head, $n_{H,head}$, and in the tail, $n_{H,tail}$.

Since equation 9 represents the self-correlation of lipid tails the variable $\mathcal{A}(Q)$ corresponds to elastic incoherent structure factor (EISF) usually determined from quasielastic neutron scattering (QENS). From a theoretical point of view $\mathcal{A}(Q)$ and EISF should allow to track a motion by NSE and QENS.[36, 41, 42] Below we test this critically by comparing the results of NSE and QENS studies.

We utilize the advantage that for simple cases closed equations exist, e.g., for a particle diffusing in a sphere, $\mathcal{A}(Q) = \left[\frac{3j_1(QR)}{(QR)}\right]^2 = \frac{9}{(QR)^6}(\sin(QR) - QR\cos(QR))^2$.[43] Here, $j_1$ is the first order spherical Bessel function and $R$ is the radius of the sphere that confines the motion of the particle. This approach is very common for QENS and has been successfully used for polymers with side-chains that have a similar number of carbons like lipid tails.[35] The crowded environment within the bilayer may impose a constraint which can be better described by a cylinder symmetry. By considering the lateral, $A_0(Q_Z) = \left[\frac{j_0(QR_L\cos(\theta))}{(QR_L\cos(\theta))}\right]^2$, and perpendicular diffusion, $B_0^0(Q_\perp) = \left[\frac{3j_1(QL\sin(\theta))}{(QL\sin(\theta))}\right]^2 \frac{1}{2}\int_0^\pi \sin(\theta)d\theta$, we obtain, $\mathcal{A}(Q) = A_0(Q_Z)B_0^0(Q_\perp)$.[44] Here, $j_0$ is the zeroth order spherical Bessel function, whereas, $R_L$ and $L$ are the radius and length of the cylinder, respectively.

### 3.1.6  Intermediate scattering function of all contributions

In summary, the dynamics of liposomes studied by NSE includes diffusion, membrane fluctuations, and confined motion. By inserting equations 4, 5, and 9, in equation 3 we obtain:

$$S_{liposome}(Q,t) = \left(n_{H,head} + n_{H,tail}\left(\mathcal{A}(Q) + (1-\mathcal{A}(Q))\exp\left(-\left(\frac{t}{\tau}\right)^\beta\right)\right)\right)\exp\left(-(\Gamma_Q t)^{2/3}\right)\exp(-D_t Q^2 t) \times S_{thickness}(Q,t) \quad (10)$$

Having identified the motion of the head groups, the tail dynamics can be analyzed more in detail, using protonated samples. Our results have shown that the contribution of $S_{thickness}(Q,t)$ appears to be negligible in fully protonated liposomes.

At the first glance with increasing the complexity of the models we seem to introduce more degrees of freedom. However, we combine several independent experimental techniques to acquire the results independently, which reduces the number of free parameters substantially. For example, we use DLS to determine the translational diffusion coefficient of the liposome, which avoids free parameters in the analysis of NSE data. In addition, we have well separated time and length scale contributions, which allow a simultaneous fit. Additionally, we include the isotopic sensitivity of neutrons to independently determine the different contributions to equation 3.

In a first step towards the understanding of the molecular dynamics in liposomes, we analyze NSE experiments on partially deuterated lipids, in which the fatty acids were contrast matched by the solvent. Suppressing the signal of the tails, confirms the importance of the tail motion in case of fully protonated samples. The following considerations improve the discussion by Zilman-Granek, because it generalizes their statement of the lateral motion of lipids and relates it directly to the molecular potential.

Moreover, as illustrated by equation 9, the scattered intensity in neutron scattering experiments is very sensitive to the number of protons and deuterons. In the case of fully hydrogenated lipids, all protons contribute to $S(Q,t)$. The number of protons in the tails is much greater than the number of protons in the head group. For example, in case of DOPC $N_{tail}$ = 66, and $N_{head}$ = 18, which leads to the fractions $n_{tail}$ = 0.79, and $n_{head}$ = 0.21, respectively. Contrast matching is the appropriate tool to distinguish head and tail motion. The signal from the contrast matched tails is completely suppressed and the relative fraction of protons in the tail, i.e. $n_{H,tail}$ = 0. In this case, the weighting parameters, $n_{H,head}$ and $n_{H,tail}$, reflect the presence or absence of the dynamic contribution of the lipid head and tail in the relaxation spectra.

### 3.2 Comparison of new with existing models

Hereafter, we introduce existing concepts to analyze neutron spectroscopy data and identify differences to our new approach. The comparison illustrates that the neutron scattering theory used to derive our model reduces the number of free parameters and provides a better understanding of their physical meaning than semi empirical concepts.

### 3.2.1 Zilman-Granek model

At first glance, equation 3 is very similar to the ansatz by Zilman-Granek (equation 5). As explained in section 3.1.2, ZG introduced a separation ansatz to include translational diffusion of the entire vesicle, in-plane lateral motion, and the height-height correlations describing the dynamics in a plane perpendicular to the flat membrane surface. Below, we show the importance of translational diffusion for our analysis and compare it with theoretical assumptions by Zilman-Granek. Unlike the approach by Zilman-Granek, we use the term $S_{tail}(Q,t)$ to describe the localized motion of lipids, without limiting it to lateral motions. Hereafter, we further advance the equation and generalize this contribution, which finally leads to a more detailed understanding of the respective correlation function.

### 3.2.2 Milner-Safran (MS) model

The Milner-Safran (MS) model has been successfully applied to analyze the membrane dynamics, such as small liposomes.[45] The MS model decomposes membrane undulations in spherical harmonics to determine shape fluctuations of microemulsion droplets.[46, 47]

$$S_{MS}(Q,t) \approx \exp(-D_t Q^2 t) \left[ 4\pi j_0^2(QR) + \sum_l F_l \times \langle u_{l0}(t) u_{l0}(0) \rangle \right] \qquad (11)$$

Here, $F_l(z)$, is the weighting factor for the autocorrelation function, $\langle u_{l0}(t) u_{l0}(0) \rangle$, with $F_l(z) = (2l+1)[(l+2)j_l(z) - z j_{l+1}(z)]^2$, and, $j$ is the Bessel functions of order $l$ and $l+1$. The idea behind this factorization is that each bending mode, $l$, contributes to $S(Q,t)$.

Similarly, to our approach the MS model uses a product ansatz and includes the translational diffusion. However the MS model takes into account only the undulation for the length scale of the particle unlike the ZG prediction used in our model that results from the integration over all undulation wave vectors between the length scale of the particle and the lower cut-off molecular length scale.[48]

While the MS model was successful in describing the dynamics of small microemulsion droplets for sizes on the order of 5 nm,[53-5] it seems to fail for vesicles of radii > 20 nm.[48, 49] Therefore, our model includes the ZG approach that yields more plausible values for bending rigidities. Our model clearly shows the importance of the contribution of the tail dynamics to the total scattering function.

### 3.2.3 Summation approach

The literature often uses a summation approach to analyze the dynamics of the liposomes[45]

$$S_{sum,1}(Q,t) = \exp(-D_t Q^2 t) \left\{ A + (1-A) \exp\left[-(\Gamma_Q t)^{2/3}\right] \right\} \quad (12)$$

A frequently used variation is the approximation by[45]

$$S_{sum,2}(Q,t) \approx A \exp(-D_t Q^2 t) + (1-A) \exp\left[-(\Gamma_Q t)^{2/3}\right] \quad (13)$$

While equation 12 is again a product ansatz that assumes independence of diffusion and membrane undulation, equation 13 is a weighted addition that includes a potential correlation between both processes. From existing experimental work it is known that both equations 12 and 13 can successfully describe experimental data and yield plausible results.[45, 50] In fact, the experimentally determined values agree within experimental accuracy.[7, 25]

At first glance, $\mathcal{A}(Q)$ used in our equation 10 and $A$ used in the literature equation 12 seem to have the same meaning. However, the literature equation 12 exclusively relates to the bending elasticity while our model describes the confined motion of the tail and the head. In this context, the literature equation 12 misses the tail motion and the parameter $A$ is an empirical parameter.

### 3.2.4 Hybrid approach

The hybrid approach was used to understand the relation between membrane bending and local reorganization of the bilayer material undergoing intermonolayer sliding.[49] The hybrid model assumes a coupling between membrane undulation as described by ZG type exponential function and an elastic contribution described by an exponential decay. The translational diffusion of the liposome is considered to be statistically independent from these two processes, which leads to[49, 51]

$$S_{hybrid}(Q,t) \approx \exp(-D_t Q^2 t)\{A_T(Q,R) + (1 - A_T(Q,R))[a_{bend} \exp(-(\Gamma_{bend} t)^{2/3}) + a_{hyb} S_{hyb}(Q,t)]\} \quad (14)$$

Where $A_T(Q,R) = 4\pi[j_o(Q,R)]^2$, with $j_0$, the zeroth-order spherical Bessel's function, and $\Gamma_Q = \Gamma_{ZG}$, the Zilman-Granek relaxation rate. The internal mode is given by $A_{int} = 1 - A_T(Q,R)$. For a rigid membrane, $S_{hyb}(Q,t) = 1$, and for highly elastic membrane the hybrid mode is given by a single exponential decay $S_{hyb}(Q,t) = \exp(-\Gamma_{hyb} t)$.

The model can describe the experimental data reasonably well for rigid membranes, however, it fails for elastic membranes.[49] The model predicts a systematic faster relaxation at longer times than that was observed experimentally.[49]

Again, it shares the similarity of statistical independence of the diffusion from undulation like our model. Unlike our model the prefactor $A_T(Q,R)$ is only related to the undulation of the membranes but not to the tail motions.

### 3.3 Comparison of the new model with experimental data

The intermediate scattering function, $S(Q,t)$, from NSE for h-DOPC, h-DMPC and h-SoyPC in D$_2$O are presented in **Figure 1**. The abbreviation for the different samples investigated is reported in Table 1. The NSE data covers a maximum $Q$-range from 0.04 to 0.16 Å$^{-1}$. The solid lines in **Figure 1** illustrate a comparison between the height correlation as defined by the ZG model, $S_{height}(Q,t)$, (**Figure 1 (a-c)**, equation 5) with our new model using the factorization approach (**Figure 1 (d-e)**, equation 10). In the fitting routine, the relaxation amplitude in equation 5 is kept as a free parameter rather than fixing it to $A = 1$. The reason for this procedure will become obvious below.

We note that the calculated $S_{height}(Q,t)$ shows deviations for short Fourier times (t < 5 ns (h-DOPC), t < 3 ns (h-DMPC) and at t < 10 ns (h-SoyPC)), even more pronounced at higher momentum transfers, $Q$'s. First, we tested whether translational diffusion can be responsible for these deviations. Following the estimates by Zilman-Granek, the effect of translational diffusion should be negligible for $t \ll \eta R^3/\kappa = 4.4$ μs. We calculated the numerical value using a radius of liposome (DOPC), $R \approx 66$ nm in D$_2$O with viscosity, $\eta_{D2O} = 1.25$ mPa·s, and $\kappa = 20$ k$_B$T.[25] At first glance, it appears that the diffusion is irrelevant and not visible in the NSE experiments. However, in a recent publication, we illustrated that translational diffusion of the liposomes can affect $S(Q,t)$ at higher Fourier times but noteworthy for $t \ll 1$ μs.[25] For this test, we used the diffusion coefficient independently determined by dynamic light scattering. We conclude that only the contribution of translational diffusion cannot explain the deviations at low Fourier times.

Therefore, we tested the influence of the confined motion. The model calculations with equation 10 describe the experimental data very well, including lower Fourier times. In the data modelling the fraction of the relative fractions of protons in the head is kept fixed to, $n_{H,head} = 0.21$, for h-DOPC, $n_{H,head} = 0.23$ h-SoyPC, and, $n_{H,head} = 0.25$ for h-DMPC. As experimentally explored by Nagao *et al.* the head group correlations hidden in the intermediate scattering function of fully protonated liposomes and can only be visualized studying partially deuterated lipids.[6, 37] Following their findings, it seems to be justified to neglect $S_{thickness}(Q,t)$ in the analysis of fully protonated liposomes. If added, this term does not visibly affect the calculated $S(Q,t)$ of fully protonated liposomes.

Table 1: *Summary of abbreviations of different phospholipids mentioned in this paper*

| *Abbreviations* | Lipid mass fraction in D$_2$O | Sample names |
|---|---|---|
| *h-DOPC* | 5 wt% | Protonated-1,2-dioleoyl-sn-glycero-phosphocholine |
| *h-DMPC* | 5 wt% | Protonated-1,2-dimyristoyl-sn-glycero-3-phosphocholine |
| *h-DSPC* | 5 wt% | Protonated-1,2-distearoyl-sn-glycero-3-phosphocholine |
| *h-SoyPC* | 5 wt% | Protonated-L-α-Phosphatidylcholine |
| *dt -DPPC* | 10 wt% | Tail deuterated-1,2-dipalmitoyl-sn-glycero-phosphocholine |
| *dt -DMPC/DSPC* | 10 wt% | Mixture of Tail deuterated-DMPC (41.5 wt%) and DSPC (48.3 wt%) in h-DMPC (4.49 wt%)-h-DSPC (1.1 wt%) |

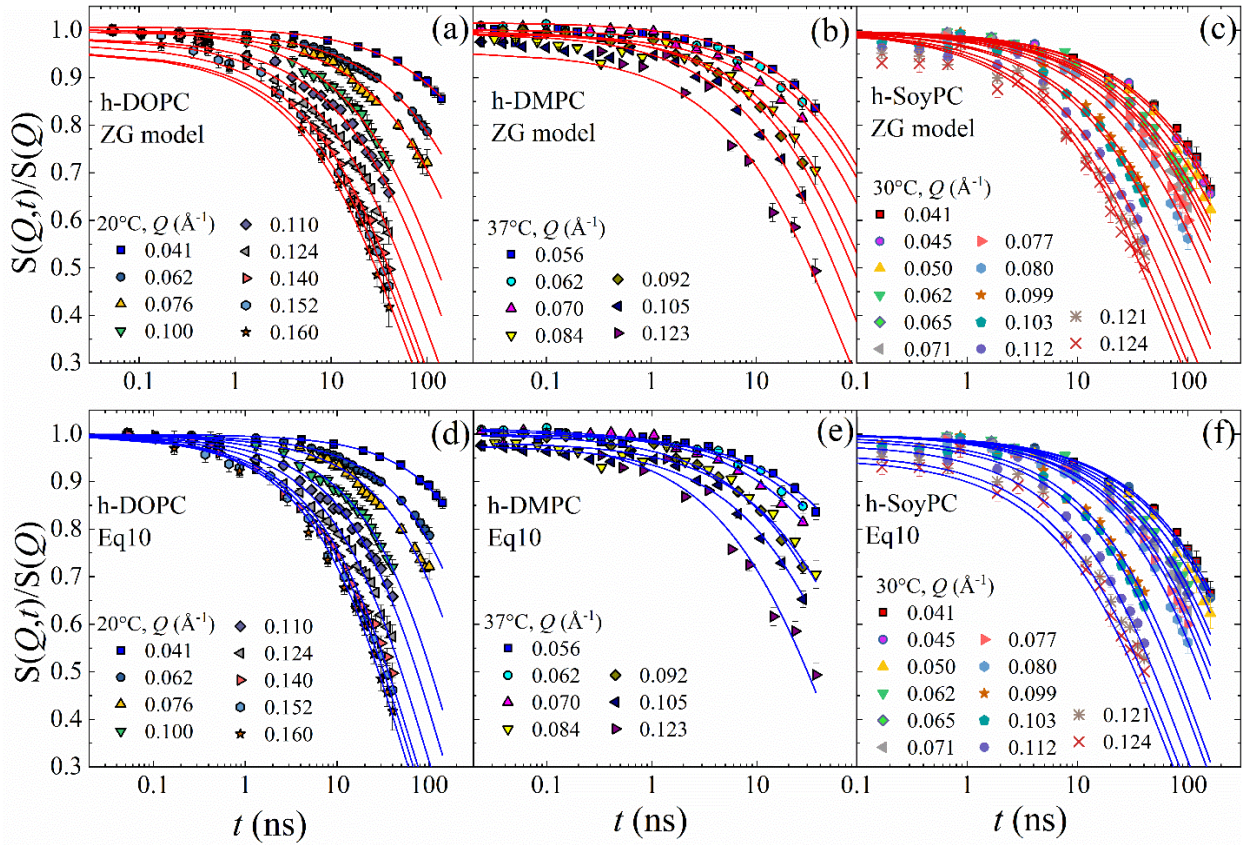

*Figure 1: Lin-log representations of the normalized intermediate scattering function, S(Q,t)/S(Q), as a function of Fourier time, t, for different Q's, for, (a, d) 5 % lipid mass fraction of protonated DOPC at 20 °C (data from reference [25]), (b, e) 5 % lipid mass fraction of protonated DMPC at 37 °C (data from reference [8]) and (c, f) the 5 % lipid mass fraction of protonated Soy-PC sample at 30 °C (data from reference [25]), each dispersed in D₂O. The same data sets are analyzed by fits using the (a-c) Zilman-Granek model (ZG) (equation 5) and (d-f) the full model that starts from equation 3 and includes diffusion and confined motion (equation 10). The error bars representing one standard deviation. The corresponding figure in log-log is presented in the electronic supplementary information.*

The NSE data illustrating the intermediate scattering function $S(Q,t)$ for tail contrast matched samples are presented in **Figure 2 (a)** and **(b)** for DPPC and for a DMPC - DSPC binary mixture, respectively.[6, 20] In these partially deuterated samples the neutron scattering length density of the tail is contrast matched with D₂O. For this case, $n_{H,head} = 1$ and $n_{H,tail} = 0$, i.e., the contribution of the tails to the intermediate scattering function in equation 10 is expected to disappear. As **Figure 2 (a)** and **(b)** illustrate the model describes the experimental $S(Q,t)$ very well. This indicates the absence of the short time contribution to the signal and connects the short-time dynamics observed in the fully protonated lipids with the motion of the fatty acid tails.

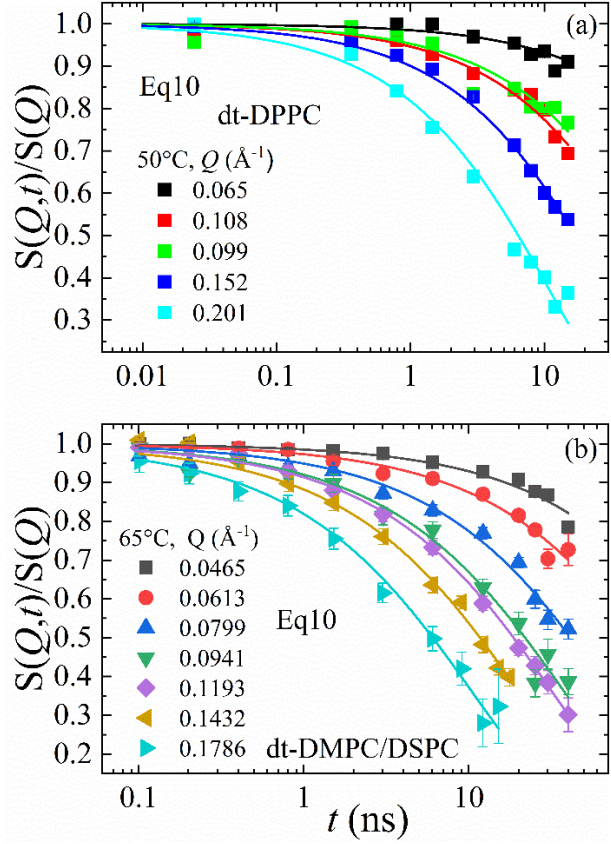

*Figure 2: Normalized intermediate scattering function, $S(Q,t)/S(Q)$, as a function of Fourier time, t, for different Q's (a) for mixture of protonated and deuterated tail DPPC in $D_2O$ sample at 50°C and (b) for the 100 mg/ml of equimolar mixture of tail contrast matched deuterated (dt) DMPC and DSPC at 65°C, each 10% lipid mass fraction. The data is fitted using our model, equation 10, with $n_{H,head}$ = 1, and, $n_{H,tail}$ = 0. NSE data are adapted from literature [6, 20].*

Apparently, data in **Figure 1** and **2** can be well described by the modeling concept. Hereafter, we use the MSD to illustrate the different contributions. Using the cumulant expansion in equation 2 and superimposing the MSDs in the ZG regime we obtain $\langle \Delta r(t)^2 \rangle_N$. The results are illustrated in Figure 3 (a) and compared with different phospholipid samples such as h-DOPC, h-DSPC, h-DMPC, and h-SoyPC.[25] The results from MD simulations of h-POPE (palmitoyl-oleoyl-phosphatidylethanolamine) are also included (grey circles).[52]

Figure 3 (a) clearly illustrates the absence of $t^{0.26}$ regime for the calculated MSDs from lipids with contrast matched tails, dt-DPPC and dt-DMPC/DSPC mixture (open circles). This does not imply the absence of the process in these samples, but rather reflects hiding the contribution of the tails for neutrons by contrast matching. More importantly, it shows the universal height-height correlation in pure lipids and lipid mixtures. It experimentally connects the emergence of the $t^{0.26}$ regime with the dynamics of the fatty acid tails. It demonstrates that if the lipid tail is invisible to the neutrons the ZG region extends to smaller Fourier times and covers the entire time window, as one observes in the analysis of single membrane layers, e.g. from microemulsions.[25] The absence of the $t^{0.26}$ adds further evidence to the argument on the hidden lipid tail motion in tail contrast matched samples. We have incorporated the relaxation spectra from equation 8 to calculate the effective MSD similar to the cumulant analysis in equation 1 and have included that in **Figure 3 (a)** for comparison. They are illustrated by the black and green solid lines for dt-DPPC and dt-DMPC/DSPC, respectively. It describes the impact of membrane thickness fluctuations on the NSE data for the tail contrast matched samples (dt-lipids).[6, 20, 37] It overlaps with the experimental data (open circles), where the deviation at t < 10 ns is missing.

The corresponding non-Gaussianity, $\alpha_2(t)$, is presented in **Figure 3 (b)**. For all protonated samples, we observe finite non-Gaussianity, $\alpha_2(t) > 0$ for low Fourier time. If the tail is contrast matched, we obtain $\alpha_2(t) = 0$ for the full-time window of our NSE experiment. This elucidates the fact that non-Gaussianity is directly related to the motion of the tail groups.

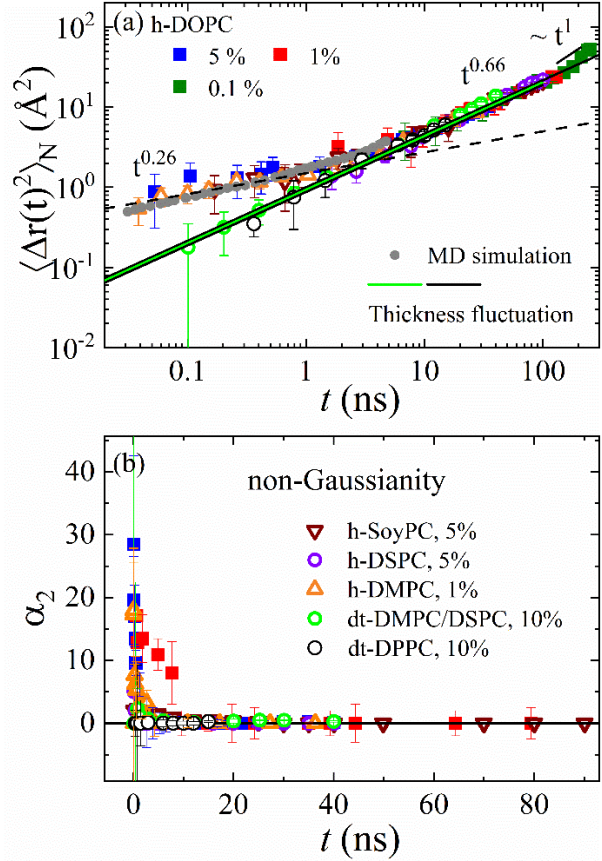

*Figure 3 (a) Normalized mean square displacement, $\langle \Delta r(t)^2 \rangle_N$, vs. Fourier time, t, for 0.1%, 1% and 5% h-DOPC, 5% h-DSPC, 1% h-DMPC and 5% h-SoyPC samples, adopted from our previous study.[25] The data for 10% dt-DMPC/DSPC mixture and 10% dt-DPPC are calculated using $S(Q,t)/S(Q)$ from the literature.[6, 20] The dashed lines represent the experimental power-law dependence, filled circles from MD simulation for h-POPE.[52] The solid lines represents the calculation for thickness fluctuation from equation 8 for dt-DPPC (black) and dt-DMPC/DSPC (green), as explained in the text. (b) The corresponding non-Gaussian parameter $\alpha_2$.*

The representation of $S(Q,t)$ by $\langle r^2(t) \rangle$ and its power-law dependence, $\langle \Delta r(t)^2 \rangle \propto t^x$, (x = 0.26 or 0.66) emphasize the fact that at least three different processes contribute to the relaxation within the length and time scale of the NSE experiments. The absence of the $t^{0.26}$ power-law for tail contrast matched samples is a direct experimental evidence that the associated $S(Q, t)$ is only connected to the dynamics of the fatty acid tails. The appearance of three different regions in $\langle r^2(t) \rangle$ emphasizes the importance to analyze the data with a function that goes beyond the simple height-height correlation model traditionally used in the literature.

With the experimental evidence of the existence of the fast-local tail motion that determines the fast relaxation we can analyze the experimental results more in detail. In a next step we will explore the motion of the tail group more in detail and obtained $\mathcal{A}(Q)$ as obtained from the fit of the experimental data by equation 10. We also compared $\mathcal{A}(Q)$ or the equivalent EISF from the QENS data.[8]

**Figure 4** (a) presents the $\mathcal{A}(Q)$ as obtained from NSE. We modeled the data by a sphere and by a cylinder. The fit values are listed in *Table 2*. However, only a dynamic Guinier plateau is visible in our NSE data. This is to be expected, because the bilayer thickness fluctuations correspond to $Q_0 \approx 0.091$ Å$^{-1}$. From this value we estimate a dynamic length $2\pi/Q_0 = 69$ Å.[6, 20] Equation 3 assumes the motion of a single lipid tail, which is less than half of the distance between the heads in the inner and outer leaflets. In other words, $Q_0$ at least doubles, which indicates that our NSE experiments did not reach the dynamic Porod region or even the transition to the dynamic Porod region. The appropriate length-scales are accessible by QENS experiments, which easily access $Q > 0.2$ Å$^{-1}$. Therefore **Figure 4** (b) includes the equivalent EISF as obtained from QENS data.

The data in **Figure 4** (a) is modeled using the $\mathcal{A}(Q)$ for a particle confined in a sphere and for a cylinder as explained in section 3.1.5. Both equally well describe the experimental results. The corresponding fit parameters are reported in **Table 2**. It should be noted that for some of the samples where the radius is less than equal to the length of the cylinder, a motion confined to a cylindrical potential could also be represented by an ellipsoidal symmetry. However, our experimental results do not permit to make such a detailed analysis.

Assuming a cylinder and realizing that the crossover to the diameter is far outside the NSE $Q$ range, we can only determine the length of the cylinder, to be between 1.4 Å and 2.7 Å for the different lipids, whereas, the length of the individual lipid molecule, $\delta_T/2$, is between 11 Å and 21 Å (**Table 2**). This comparison indicates that the confinement is caused within ~ $1/8^{th}$ the length of the lipid tail, which is approximately the size of the $CH_2$ or $CH_3$ part of the acyl group of the fatty acid.[53]

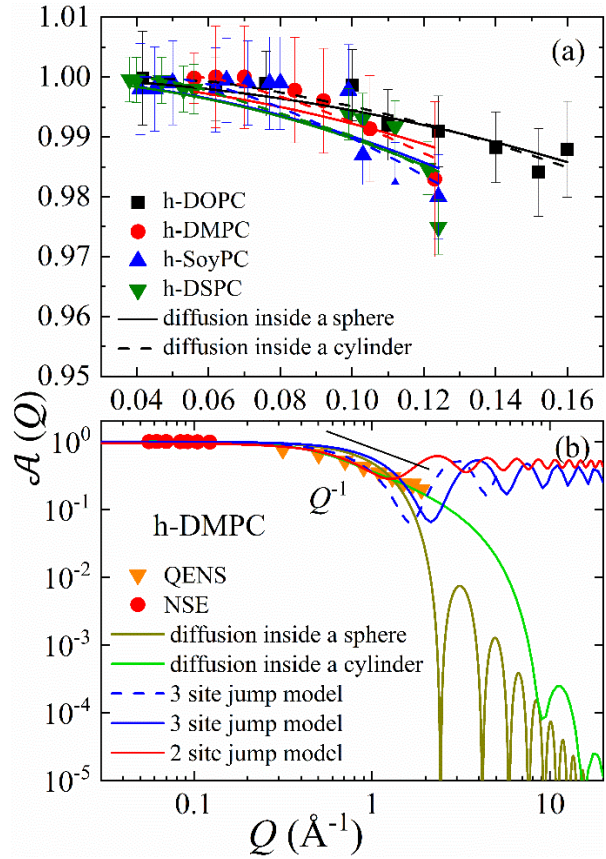

*Figure 4: (a) The $\mathcal{A}(Q)$ obtained from modeling the NSE relaxation spectra following equation 9. The solid and dashed lines are fits using the EISF for a particle diffusion in a sphere and cylinder models, respectively. (b) The $\mathcal{A}(Q)$ for h-DMPC obtained from NSE and QENS studies,[8] over a broad Q-range. The data is modeled using $\mathcal{A}(Q)$ for a sphere, cylinder in comparison with three and two site jump models. The error bars representing one standard deviation. The two-site jump model with a radius of 1.5 Å (solid red line) is compared with three-site jump model for a radius of 1.34 Å (solid blue line) and 0.99 Å (dashed blue line).*

To extend the length ($Q$-range) and time scale of the observed dynamic confinement in **Figure 4 (b)** we have included the $\mathcal{A}(Q)$ obtained from quasi-elastic neutron scattering (QENS) experiments.[8] The data from NSE and QENS are modeled simultaneously.

The fatty acid tails are mobile objects. Thus, several processes could account for $\mathcal{A}(Q)$. A spherical potential, a lipid confined to a cylinder, a two-site jump model of the lipid tails, which is related to rotational diffusion of the head perpendicular to the bilayer, and three site jumps of the protons in the methyl group. The lipid molecule has a total of 5 methyl groups, with 2 in the tails and 3 in the head group that can contribute to the signal. The results are displayed in **Figure 4 (b)**, the fitting values are listed in **Table 2**.

We can describe the experimental data by a two-site jump model choosing a radius of 1.5 Å (solid red line), whereas the three-site jump model is calculated for using 1.34 Å (solid blue line) and 0.99 Å (dashed blue line). The last value represent the distance from each H-atom of a methyl group to the center of gravity is 0.99 Å.[41] These are the values where we find the closest match to the experimental results. However, we witness notable discrepancies. Therefore, despite the existence of these motions their contribution does not strongly affect the experimental data.

The diffusion inside a cylinder with length $L = 3.72 \pm 0.2$ Å and radius $R_L$ set to 0.5 Å yields the best description. From the fit of the dynamic Guinier range alone, we obtain $L = 3.73 \pm 0.4$ Å. These values are very close to an independent QENS study on h-DMPC by Wanderlingh *et al.*[54] who report $L = 3.73$ Å and $R_L = 4.25$ Å. The diameter of the cylinder is very close to the distance between two $CH_3$ groups in the fatty acid tail. However, we note that these values are only an estimate, because even the QENS experiment does not resolve the dynamic Porod region.

Table 2: *Summary of the lipid tail motion considering a potential of spherical symmetry of radius, R, or a cylindrical object of radius, $R_L$, and length, L, obtained from the analysis of the data in the* **Figure 4** *by equation 10. The lipid tail thickness, $\delta_T$, from literature, and the estimates of the relaxation time, τ, of the confined tail is reported. The gel-fluid transition temperature, $T_m$,[27] and the distance to the measurement temperature, $T - T_m$, from the literature illustrates that all samples are in the fluid state.*

| Samples | $T_m$ (°C) | $T - T_m$ (°C) | A(Q), Sphere, R (Å) | A(Q), Cylinder | | Lipid tail thickness $\delta_T$ (Å) (literature) | τ (ns) |
|---|---|---|---|---|---|---|---|
| | | | | $R_L$ (Å) | L (Å) | | |
| h-DOPC | -16.5 | 36.5 | 1.7 ± 0.1 | 2 | 1.4 ± 0.1 | 25.00 ± 0.05[25] | 2.8 |
| h-DMPC | 23.6 | 13.4 | 2.0 ± 0.2 | 0.5 | 3.7 ± 0.2 | 22.6 ± 0.6[6] | 2.0 |
| h-DSPC | 54.4 | 10.6 | 2.3 ± 0.1 | 2 | 1.9 ± 0.2 | 32 ± 0.2[6] | 3.0 |
| h-SoyPC | -18.5 | 48.5 | 2.2 ± 0.2 | 3 | 2.1 ± 0.2 | 23 ± 3[55] | 3.2 |
| dt-DPPC | 37.5 | 12.6 | N/A | N/A | N/A | 30 ± 0.3[6] | N/A |
| dt-DMPC/DSPC | 20.5 / 50.5 | 44.5 / 14.5 | N/A | N/A | N/A | 40.9 ± 10[6, 20] | N/A |

The length of the fully extended tail of h-DMPC is between, 11 Å and 13 Å ($\delta_T/2$ in **Table 2**), our observed length of the cylinder ~ 1/3$^{rd}$ of that. This indicates a strong confinement inside the lipid bilayer. It should be noted that all these length scales correspond to a dynamic confinement length, rather than the static lengths. The dynamic length of a lipid is not expected to match the static value. However, in this case, the well-fitted data from NSE and QENS confirm our assumption that we can model $\mathcal{A}(Q)$ from NSE and QENS for the lipid tail motion simultaneously.

The importance of the spherical confinement for the lipid motion has been extensively studied using QENS. Previous QENS study have revealed the existence of solvation cage for the whole lipid molecule in the fluid phase,[56] whereas, the motion of the lipid tail is highly heterogeneous.[57] It was also suggested in combination of MD simulations and QENS that this dynamic heterogeneity originates from the fact that in a spherical confinement the proton diffusion is greater at the chain ends than at the glycerol backbone.[57, 58]

Table 3: *Bending moduli, $\kappa_\eta$ as obtained by the analysis of the data in* **Figure 5** *by equation 7, and from literature. Please note that some of the values from literature required a recalculation to account for different numerical prefactors used in the literature. For the calculation of $\kappa_\eta$ we used the prefactor 0.0069 as detailed in section 3.1.3, cf. equation 6. The ratio $\kappa_\eta$(literature) /$\kappa_\eta$ is included for comparison. The gel-fluid transition temperature, $T_m$,[27] and the distance to the measurement temperature, $T - T_m$, from the literature illustrates that all samples are in the fluid state.*

| Samples | $T_m$ (°C) | $T - T_m$ (°C) | $\kappa_\eta/k_B T$ | $\kappa_\eta/k_B T$ (literature) | Ratio $\kappa_\eta$(literature) /$\kappa_\eta$ |
|---|---|---|---|---|---|
| h-DOPC | -16.5 | 36.5 | 18 ± 2 | 23 ± 1 [7] | 1.3 |
| h-DMPC | 23.6 | 13.4 | 12 ± 3 | 24.6 ± 1.3 [8] | 2.1 |
| h-DSPC | 54.4 | 10.6 | 23 ± 3 | 42.0 ± 1.2 [37] | 1.8 |
| h-SoyPC | -18.5 | 48.5 | 6.0 ± 2 | 8.4 ± 1 [25] | 1.4 |
| dt -DPPC | 37.5 | 12.6 | 19.5 ± 2 | 24.2 ± 2 [6] | 1.2 |
| dt -DMPC/DSPC | 20.5 / 50.5 | 44.5 / 14.5 | 13 ± 2 | 28.0 ± 1 [20] | 2.2 |

We note an obvious difference to bicontinuous microemulsions in which diffusion is absent. There, $\langle r^2(t) \rangle \propto t^{0.66}$ which indicates that only height-height correlations can be found. Thus, the analysis by the ZG model, or the asymptotic approach,[59] or the more sophisticated MS model[47] is valid. On the other hand, it becomes clear that our results indicate that the analysis by a simple ZG model (without considering additional effects) is not enough and necessarily leads to inaccuracies in the parameters. Since the ZG model is very common in the literature, we now attempt to estimate the errors involved in neglecting the local lipid motion.

For that purpose, we use equation 7 to determine the bending rigidity, $\kappa_\eta/k_B T$, as a function of the Fourier time from $\langle \Delta r(t)^2 \rangle_N$ in **Figure 3.** The results are illustrated in **Figure 5**. It is obvious that $\kappa_\eta$ has a pronounced time dependence, initially proportional to $t^{1.22 \pm 0.09}$, for $\kappa_\eta/k_B T \propto t^{2-3x}$, $x = 0.26 \pm 0.03$. The constant full lines represent the expectations from the ZG model, $t^0$. We included those values from the analysis of our data by the ZG model and added the bending rigidities determined from the multiplicative approach (equation 10).

At the first glance even the more advanced model seems to have some discrepancies with the experimental data. However, this is related to the fact, that the calculated $\kappa_\eta$ is affected by all motions, including the translational diffusion.

One can expect a constant value for $\kappa_\eta/k_B T$ over the calculated time window. However, the strong deviation from the constant value at t < 5 ns is a result of the finite non-Gaussianity, $\alpha_2(t) \neq 0$. The average value of $\kappa_\eta$ in the ZG regime is presented in **Table 3**. The deviation from the $t^0$ prediction of the ZG model suggests presence of additional dynamics.[60, 61]

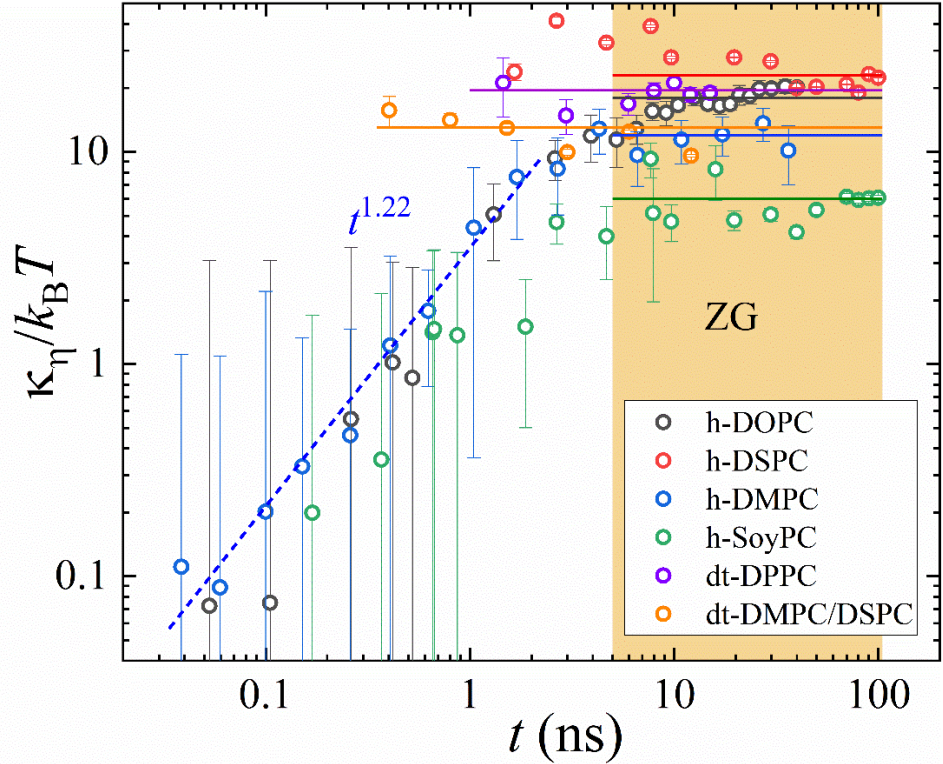

*Figure 5: The membrane rigidity calculated over the entire NSE time window from the MSD using equation 7. The results for protonated and partially deuterated lipids are presented for comparison. The error bars represent one standard deviation in a log-log plot. The NSE data for DMPC, DPPC and DSPC are calculated using $S(Q,t)$ from the literature.[6, 8, 20] The NSE data for DOPC, Soy-PC, DSPC are from our previous study.[25] A comparison to our calculated $t^{1.22\pm0.09}$ power-law dependence is illustrated by the dashed line.*

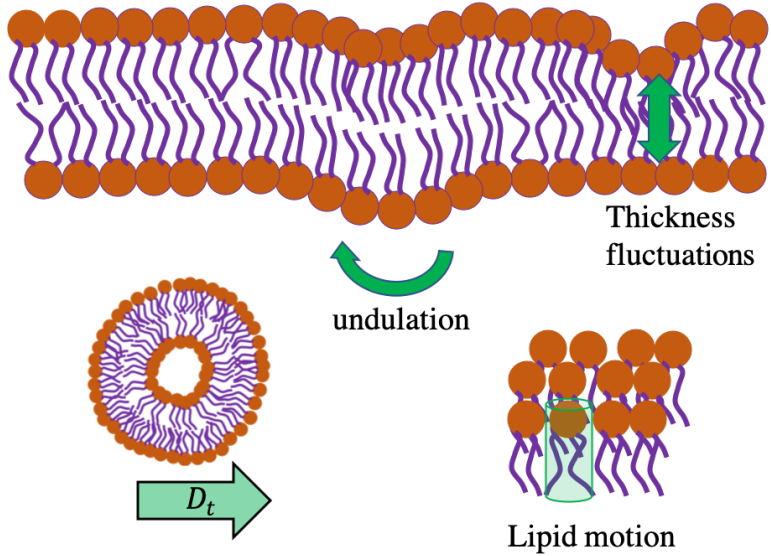

*Figure 6: Schematic representation of the different dynamics of the liposome and the lipid bilayer as discussed in this paper.*

## 4 CONCLUSION

We presented experimental evidence of the existence of constrained local dynamics inside the lipid bilayer using neutron spin echo spectroscopy (NSE). A comparison of the MSD from fully protonated and tail contrast matched phospholipids reveals the absence of the $t^{0.26}$ power law in tail contrast matched samples. Experimental result and analysis relate the fast time dynamics very strongly to the motion of the lipid tails. Our results demonstrate that the time-scales for the fast-local relaxation of lipids and the height-height correlation of membranes can be treated by statistically independent functions, which clearly shows the need for the new model function derived in the present work. We demonstrated the limitation of the ZG model to a finite time range between a fast and a slow motion, i.e., time range approximately from 5 to 100 ns. The slow motion was identified to be the translational diffusion of liposomes. If not included then the overall relaxation behavior is not analyzed correctly, especially at long Fourier times. The analysis of the fast dynamics connects the dynamics of the lipid tails with a very confined motion. It cannot be described by the ZG model that assumes height-height correlations. Independently of its origin it needs to be included in the considerations, otherwise the fit provides wrong values for the bending elasticity. Furthermore, our results demonstrate that the need of a better understanding of neutron spectroscopic data, e.g., by including parameters like the translation diffusion of liposomes from dynamic light scattering. For example, if the time range of the NSE experiment is too limited, then DLS is the only means to determine the most accurate value, but NSE can utilize it to improve the accuracy of the result on the bending elasticity. A schematic illustration of the different dynamics is presented in **Figure 6**.

The simplest model that is compatible with our data at fast Fourier times is a potential with cylindrical symmetry. Our analysis emphasizes the importance of the motion of the lipid tails over a broad range of length-scales. The present paper advances the understanding, by relating the term trapped motion to confined motion. This is the first experimental evidence that identifies the origin and the nature of the trapped motion in the bilayer over multiple length and time scale.

The availability of experimental data over a broad range could advance older literature, e.g., in which the confined motion of lipids was described by a spherical potential using a distribution of confinement sizes.[57] In other words, the results strongly indicate that the lipids relax in a cylindrical confinement, where the dynamic length scale represents only around about 1/3$^{rd}$ the length of the lipid tail.

The MSD shows power laws $t^n$ with $n < 1$. These so-called sub-diffusive motions are assumed to be important for cellular signaling and regulatory process. Transient trapping or the confined motion has a power law with $n = 0.26$. Numerous examples connect transient trapping to biophysical processes. (i) It has been reported that it is important for compartmentalization of mRNA into smaller subcellular regions in living cells.[62] Clustering of "gene encoding interacting proteins" in this confined space facilitates a transfer of genetic information between living cells. (ii) It has been shown that the length scale associated with transient trapping corresponds to the distance that proteins move to find binding sites on DNA.[63] (iii) A similar phenomenon has also been observed for transmembrane proteins that recognize specific adaptor molecules for binding.[64] (iv) Recent studies on potassium channels of the plasma membrane of living cells have demonstrated the anomalous nature of the diffusion following a transient trap defined by CTRW model described by the observed non-Gaussianity.[15]

It should be noted that following the CTRW model by Akimoto *et al*.[52] the importance of dynamic heterogeneity behind the origin of transient trapping of the lipid tail, where the lipid tail in the fluid phase are disordered and randomly oriented, similar to that observed in colloids[65] and glassy materials.[66] The ability to identify the confined motion in experimental data, to analyze it and to study the impact of different environments is important and stimulates future studies.

**AUTHOR INFORMATION**
**Corresponding Authors**

*E-mail: g.sudipta26@gmail.com

*E-mail: gjschneider@lsu.edu

**ORCID**

Sudipta Gupta: 0000-0001-6642-3776

Gerald J. Schneider: 0000-0002-5577-9328

**Notes**

The authors declare no competing financial interest.

**ACKNOWLEDGEMENT**

The neutron scattering work is supported by the U.S. Department of Energy (DOE) under EPSCoR Grant No. DE-SC0012432 with additional support from the Louisiana Board of Regents. This paper was prepared as an account of work sponsored by an agency of the United States Government. We would like to acknowledge Dr. Antonio Faraone for assisting us with the neutron spin echo spectrometer from National Institute of Standards and Technology (NIST). Access to the neutron spin echo spectrometer was provided by the Center for High Resolution Neutron Scattering, a partnership between the National Institute of Standards and Technology (NIST) and the National Science Foundation under Agreement No. DMR-1508249. We would like to acknowledge Dr. Piotr Zolnierczuk for assisting us with the neutron spin echo spectrometer from Spallation Neutron Source (SNS) at Oak Ridge National Laboratory (ORNL). Research conducted at the Spallation Neutron Source (SNS) at Oak Ridge National Laboratory (ORNL) was sponsored by the Scientific User Facilities Division, Office of Basic Energy Sciences, U.S. DOE.

# REFERENCES

1. D. Lingwood and K. Simons, *Science*, 2010, **327**, 46-50.
2. K. Simons and E. Ikonen, *Nature*, 1997, **387**, 569-572.
3. J. Katsaras and T. Gutberlet, *Lipid Bilayers: Structure and Interactions*, Springer-Verlag, Berlin, Heidelberg,, 2001.
4. L. F. Aguilar, J. A. Pino, M. A. Soto-Arriaza, F. J. Cuevas, S. Sánchez and C. P. Sotomayor, *PLOS ONE*, 2012, **7**, e40254.
5. A. K. Menon and A. Herrmann, in *Encyclopedia of Biophysics*, ed. G. C. K. Roberts, Springer Berlin Heidelberg, Berlin, Heidelberg, 2013, DOI: 10.1007/978-3-642-16712-6_551, pp. 1261-1264.
6. A. C. Woodka, P. D. Butler, L. Porcar, B. Farago and M. Nagao, *Phys Rev Lett*, 2012, **109**, 058102.
7. I. Hoffmann, R. Michel, M. Sharp, O. Holderer, M. S. Appavou, F. Polzer, B. Farago and M. Gradzielski, *Nanoscale*, 2014, **6**, 6945-6952.
8. V. K. Sharma, E. Mamontov, M. Ohl and M. Tyagi, *Phys Chem Chem Phys*, 2017, **19**, 2514-2524.
9. X. Yi, X. Shi and H. Gao, *Phys Rev Lett*, 2011, **107**, 098101.
10. L. Movileanu, D. Popescu, S. Ion and A. I. Popescu, *Bulletin of Mathematical Biology*, 2006, **68**, 1231-1255.
11. T. M. Allen and P. R. Cullis, *Advanced Drug Delivery Reviews*, 2013, **65**, 36-48.
12. O. Stauch, R. Schubert, G. Savin and W. Burchard, *Biomacromolecules*, 2002, **3**, 565-578.
13. A. Sumino, T. Dewa, T. Takeuchi, R. Sugiura, N. Sasaki, N. Misawa, R. Tero, T. Urisu, A. T. Gardiner, R. J. Cogdell, H. Hashimoto and M. Nango, *Biomacromolecules*, 2011, **12**, 2850-2858.
14. S. Block, V. P. Zhdanov and F. Hook, *Nano Lett*, 2016, **16**, 4382-4390.
15. A. V. Weigel, B. Simon, M. M. Tamkun and D. Krapf, *Proc Natl Acad Sci U S A*, 2011, **108**, 6438-6443.
16. E. C. Jensen, *Anat Rec (Hoboken)*, 2012, **295**, 2031-2036.
17. C. W. Harland, M. J. Bradley and R. Parthasarathy, *Proc Natl Acad Sci U S A*, 2010, **107**, 19146-19150.
18. I. Schmidt, F. Cousin, C. Huchon, F. Boue and M. A. Axelos, *Biomacromolecules*, 2009, **10**, 1346-1357.
19. J. D. Nickels, X. Cheng, B. Mostofian, C. Stanley, B. Lindner, F. A. Heberle, S. Perticaroli, M. Feygenson, T. Egami, R. F. Standaert, J. C. Smith, D. A. Myles, M. Ohl and J. Katsaras, *J Am Chem Soc*, 2015, **137**, 15772-15780.
20. R. Ashkar, M. Nagao, P. D. Butler, A. C. Woodka, M. K. Sen and T. Koga, *Biophys J*, 2015, **109**, 106-112.
21. H. E. Cingil, W. H. Rombouts, J. van der Gucht, M. A. Cohen Stuart and J. Sprakel, *Biomacromolecules*, 2015, **16**, 304-310.
22. T. Gibaud, C. N. Kaplan, P. Sharma, M. J. Zakhary, A. Ward, R. Oldenbourg, R. B. Meyer, R. D. Kamien, T. R. Powers and Z. Dogic, *Proc Natl Acad Sci U S A*, 2017, **114**, E3376-E3384.
23. S. Gupta, R. Biehl, C. Sill, J. Allgaier, M. Sharp, M. Ohl and D. Richter, *Macromolecules*, 2016, **49**, 1941-1949.
24. S. Xuan, S. Gupta, X. Li, M. Bleuel, G. J. Schneider and D. Zhang, *Biomacromolecules*, 2017, **18**, 951-964.
25. S. Gupta, J. U. De Mel, R. M. Perera, P. Zolnierczuk, M. Bleuel, A. Faraone and G. J. Schneider, *J Phys Chem Lett*, 2018, **9**, 2956-2960.
26. S. Gupta, J. U. De Mel and G. J. Schneider, *J Phys Chem B*, 2019, **123**, 5667-5669.
27. S. Gupta, J. U. De Mel and G. J. Schneider, *Current Opinion in Colloid & Interface Science*, 2019, **42**, 121-136.
28. A. G. Zilman and R. Granek, *Phys Rev Lett*, 1996, **77**, 4788-4791.
29. R. B. Pandey, K. L. Anderson and B. L. Farmer, *Phys Rev E Stat Nonlin Soft Matter Phys*, 2007, **75**, 061913.
30. E. Flenner, J. Das, M. C. Rheinstadter and I. Kosztin, *Phys Rev E Stat Nonlin Soft Matter Phys*, 2009, **79**, 011907.
31. J. H. Jeon, H. M. Monne, M. Javanainen and R. Metzler, *Phys Rev Lett*, 2012, **109**, 188103.
32. J.-H. Jeon, M. Javanainen, H. Martinez-Seara, R. Metzler and I. Vattulainen, *Physical Review X*, 2016, **6**, 021006.
33. R. Zorn, *Physical Review B*, 1997, **55**, 6249-6259.
34. G. J. Schneider, K. Nusser, S. Neueder, M. Brodeck, L. Willner, B. Farago, O. Holderer, W. J. Briels and D. Richter, *Soft Matter*, 2013, **9**, 4336.
35. C. Gerstl, G. J. Schneider, A. Fuxman, M. Zamponi, B. Frick, T. Seydel, M. Koza, A. C. Genix, J. Allgaier, D. Richter, J. Colmenero and A. Arbe, *Macromolecules*, 2012, **45**, 4394-4405.
36. Julia S. Higgins and Henri C. Benoit, *Polymers and Neutron Scattering*, Clarendon Press, Oxford New York, 1996.

37. M. Nagao, E. G. Kelley, R. Ashkar, R. Bradbury and P. D. Butler, *J Phys Chem Lett*, 2017, DOI: 10.1021/acs.jpclett.7b01830, 4679-4684.
38. M. C. Watson and F. L. Brown, *Biophys J*, 2010, **98**, L9-L11.
39. T. Takeda, Y. Kawabata, H. Seto, S. Komura, S. K. Ghosh, M. Nagao and D. Okuhara, *Journal of Physics and Chemistry of Solids*, 1999, **60**, 1375-1377.
40. R. J. Bingham, S. W. Smye and P. D. Olmsted, *Europhysics Letters (EPL)*, 2015, **111**, 18004.
41. M. Beé, *Quasielastic neutron scattering principles and applications in solid state chemistry, biology and materials science*, Adam Hilger, Bristol, England, 1988.
42. V. F. Sears, *Neutron Optics: An Introduction to the Theory of Neutron Optical Phenomena and their Applications*, Oxford University Press, 1989.
43. F. Volino and A. J. Dianoux, *Molecular Physics*, 2006, **41**, 271-279.
44. A. J. Dianoux, M. Pineri and F. Volino, *Molecular Physics*, 1982, **46**, 129-137.
45. L. R. Arriaga, I. Lopez-Montero, G. Orts-Gil, B. Farago, T. Hellweg and F. Monroy, *Phys Rev E Stat Nonlin Soft Matter Phys*, 2009, **80**, 031908.
46. I. R. Miller, *Biophysical Journal*, 1984, **45**, 643-644.
47. S. T. Milner and S. A. Safran, *Physical Review A*, 1987, **36**, 4371.
48. I. Hoffmann, C. Hoffmann, B. Farago, S. Prevost and M. Gradzielski, *J Chem Phys*, 2018, **148**, 104901.
49. M. Mell, L. H. Moleiro, Y. Hertle, I. Lopez-Montero, F. J. Cao, P. Fouquet, T. Hellweg and F. Monroy, *Chem Phys Lipids*, 2015, **185**, 61-77.
50. M. Mell, L. H. Moleiro, Y. Hertle, P. Fouquet, R. Schweins, I. Lopez-Montero, T. Hellweg and F. Monroy, *Eur Phys J E Soft Matter*, 2013, **36**, 75.
51. M. B. Schneider, J. T. Jenkins and W. W. Webb, *Journal de Physique*, 1984, **45**, 1457-1472.
52. T. Akimoto, E. Yamamoto, K. Yasuoka, Y. Hirano and M. Yasui, *Phys Rev Lett*, 2011, **107**, 178103.
53. J. F. Nagle and S. Tristram-Nagle, *Biochimica et Biophysica Acta (BBA) - Reviews on Biomembranes*, 2000, **1469**, 159-195.
54. U. Wanderlingh, G. D'Angelo, C. Branca, V. C. Nibali, A. Trimarchi, S. Rifici, D. Finocchiaro, C. Crupi, J. Ollivier and H. D. Middendorf, *J Chem Phys*, 2014, **140**, 174901.
55. G. Mangiapia, M. Gvaramia, L. Kuhrts, J. Teixeira, A. Koutsioubas, O. Soltwedel and H. Frielinghaus, *Phys Chem Chem Phys*, 2017, **19**, 32057-32071.
56. S. König, T. M. Bayerl, G. Coddens, D. Richter and E. Sackmann, *Biophysical Journal*, 1995, **68**, 1871-1880.
57. S. König, W. Pfeiffer, T. Bayerl, D. Richter and E. Sackmann, *Journal de Physique II France*, 1992, **2**, 1589-1615.
58. M. Doxastakis, V. G. Sakai, S. Ohtake, J. K. Maranas and J. J. de Pablo, *Biophys J*, 2007, **92**, 147-161.
59. M. Mihailescu, M. Monkenbusch, H. Endo, J. Allgaier, G. Gompper, J. Stellbrink, D. Richter, B. Jakobs, T. Sottmann and B. Farago, *The Journal of Chemical Physics*, 2001, **115**, 9563.
60. L. Wang, K. Fujimoto, N. Yoshii and S. Okazaki, *J Chem Phys*, 2016, **144**, 034903.
61. N. Yoshii, Y. Nimura, K. Fujimoto and S. Okazaki, *J Chem Phys*, 2017, **147**, 034906.
62. P. Montero Llopis, A. F. Jackson, O. Sliusarenko, I. Surovtsev, J. Heinritz, T. Emonet and C. Jacobs-Wagner, *Nature*, 2010, **466**, 77-81.
63. G. Kolesov, Z. Wunderlich, O. N. Laikova, M. S. Gelfand and L. A. Mirny, *Proc Natl Acad Sci U S A*, 2007, **104**, 13948-13953.
64. D. J. Owen, B. M. Collins and P. R. Evans, *Annu Rev Cell Dev Biol*, 2004, **20**, 153-191.
65. S. Gupta, J. Stellbrink, E. Zaccarelli, C. N. Likos, M. Camargo, P. Holmqvist, J. Allgaier, L. Willner and D. Richter, *Phys Rev Lett*, 2015, **115**, 128302.
66. S. Gupta, J. K. H. Fischer, P. Lunkenheimer, A. Loidl, E. Novak, N. Jalarvo and M. Ohl, *Scientific Reports*, 2016, **6**, 35034.